# Toward a Cohesive AI and Simulation Software Ecosystem for Scientific Innovation[1]

Michael A. Heroux[2] ParaTools, Inc.
Sameer Shende, ParaTools, Inc.
Lois Curfman McInnes, Argonne National Laboratory
Todd Gamblin, Lawrence Livermore National Laboratory
James M. Willenbring, Sandia National Laboratories

In this document, we outline key considerations for the next-generation software stack that will support scientific applications integrating AI and modeling & simulation (ModSim) to provide a unified AI/ModSim software stack.

**Key Points:**

1. The scientific computing community needs a cohesive AI/ModSim software stack.
2. This stack must ensure version compatibility across a diverse collection of community software on diverse computing systems and be updated regularly to support new libraries and tools and new versions of existing products.
3. This AI/ModSim stack must support binary distributions to enable emerging scientific workflows.
4. Investment in a unified AI/ModSim community stack that complements computer system stacks is essential.

**A Cohesive Software Stack for AI and Modeling & Simulation**

To address future scientific challenges, the next-generation scientific software stack must provide a cohesive portfolio of libraries and tools that facilitate AI and ModSim approaches. As scientific research becomes increasingly interdisciplinary, scientists require both of these toolsets to address complex, data-rich problems in problem domains such as climate modeling, material discovery, and energy optimization. A unified software ecosystem integrating established AI frameworks and emerging scientific AI frameworks

---

[1] This document was originally submitted as a Response to the Request for Information on Frontiers in AI for Science, Security, and Technology (FASST) Initiative, addressing the question: *How can DOE continue to support the development of AI hardware, algorithms, and platforms tailored for science and engineering applications in cases where the needs of those applications differ from the needs of commodity AI applications? How can DOE partner with other compute capability providers, including both on-premises and cloud solution providers, to support various hardware technologies and provide a portfolio of compute capabilities for its mission areas?*

[2] Primary contact: Michael A. Heroux, ParaTools, Inc., 18125 Kreigle Lake Road, Avon, MN 56310, mheroux@paratools.com, Phone: +1 320 905 1275

alongside established ModSim libraries is crucial for solving these next-generation scientific challenges effectively.

**Leveraging Current Efforts: DOE Office of Science Efforts in Software Stewardship**

The DOE Office of Science has sponsored significant software stewardship and advancement efforts, especially in the post-Exascale Computing Project (ECP) era, to provide long-term support for libraries and tools used by the DOE community. These efforts include funding for math, data and visualization, performance analysis, and programming systems libraries and tools, as well as for curating a portable, reliable, and performant version-compatible portfolio of these libraries and tools and their dependencies. While the current DOE-supported stack includes many AI libraries and tools that our scientific community needs, installing the comprehensive AI stack is challenging and labor-intensive, especially on high-performance computing (HPC) systems. The PESO Project, led by the authors of this document, is one of the projects involved in post-ECP efforts, sponsoring efforts to expand the use of Spack in open-science codes and to curate and deliver E4S, the software stack created by ECP.

**Version Management and Compatibility Challenges**

One critical challenge in developing such a cohesive stack is version management and compatibility across different tools and libraries. Both the AI and MS communities face significant challenges in this area, but they have taken distinct approaches to managing versions and software upgrades. In the AI community, package managers like Pip and Conda are widely used to simplify installation for individual users. While these tools are effective for deploying AI libraries quickly, they often do not adequately address broader compatibility issues, which makes integration with ModSim tools difficult. This gap results in challenges when scientists need to integrate numerous AI libraries and ensure consistent behavior across diverse computing environments. To enable smooth integration, we need a comprehensive and cohesive portfolio that includes both AI and ModSim components.

**Building Software from Source vs. Prebuilt Binaries**

On the other hand, the ModSim community has leaned toward building software from source, which has traditionally provided more flexibility for configuring and optimizing for performance. However, as the need for more productive scientific computing environments grows, the demand for ease of deployment has grown. This community also increasingly requires prebuilt binaries of core libraries and tools, leading to similar challenges faced by the AI community. Tools like CMake are often used to manage builds, and Spack has become an essential tool for managing dependencies and configurations across various libraries and platforms. Spack provides not only the capability to build from source but also to create reusable binaries, offering an alternative path for rapid software deployment. Spack and E4S both offer reusable binary caches that the community has become accustomed to.

**User Interaction with Computing Resources**

Another distinction between AI and ModSim approaches is how users interact with their computing resources. The AI community is based heavily in the Python ecosystem. PyTorch, TensorFlow, JAX, and related tools expose powerful Python front-ends to users, and users may drive computation from Jupyter notebooks, or from orchestration tools like Kubernetes, KubeFlow, or Argo. Traditional high-performance ModSim relies primarily on batch schedulers (e.g., SLURM or Flux) that do not have good support for persistent services, databases, or interactive web applications used throughout the AI community. AI models themselves require versioning, regular updates, and bugfixes, and AI inference is typically deployed *as* a persistent service. These processes are referred to in industry as "MLops". A modern AI for science software environment that combines AI and ModSim to solve complicated scientific problems.  ModSim needs to support both cloud-like persistent services (orchestration tools, notebooks, databases, etc.) and a traditional batch environment.

**Developing a Consistent and Evolving Software Stack**

Providing a consistent and up to date software stack that incorporates AI and ModSim software libraries and tools will require significant effort in identifying the core requirements, converting those requirements to specifications and design, and producing a cohesive software stack that is easily used in a portable way and available with significant updates on a regular schedule that users can count on. While present AI users are typically using industry-provided capabilities that target a broad spectrum of domains, we anticipate the creation and wide use of AI libraries and tools specifically trained for scientific problems. These libraries and tools must be curated and provided to the scientific community, especially on DOE leadership computing platforms.

**Need for a Portable and Cohesive Library and Tool Ecosystem**

Traditionally, computing system vendors have provided these kinds of libraries. However, insufficient effort has been applied to making a portable and cohesive library and tool ecosystem that users can rely on, independent of which vendor's platform they are running on. The Exascale Computing Project made progress in providing a portable, high-performance software stack that provides libraries and tools for established ModSim applications for users on major HPC platforms. The same basic approach needs to be explored for high-performance, portable software solutions on the growing variety of AI for science computing platforms. While computer system vendors such as NVIDIA and AMD are vested in providing software stacks for their users, the community can benefit from an industry effort that is not tied to any specific hardware platform. In collaboration with system vendors and the Department of Energy, independent scientific software companies can play an important role in this effort.

**A Unified Framework for AI and ModSim Software**

The complexities of managing AI and ModSim software products go beyond supporting individual libraries and tools products and highlight the need for DOE to support efforts that bring AI and ModSim libraries and tools into a integrated stack. This approach is particularly important when ensuring compatibility across different versions of programming systems (for example, different Python and C++ standards), hardware platforms, and performance requirements.  This integrated approach would allow scientists to seamlessly combine AI and ModSim capabilities, leveraging the strengths of both to address next-generation scientific problems more effectively.

**Role of Continuous Integration**

To build a portable, integrated software stack for AI and ModSim, the task of updating and integrating the software stack *cannot* be manual. Continuous Integration has become a critical part of modern software development, but DOE platforms are bespoke and sit behind significant security barriers for open-source developers. In particular, key open-source projects cannot currently test and rapidly ensure that the latest versions of their fast-moving software runs on platforms of interest to DOE.  This is a chicken/egg problem– adoption for new GPUs in the cloud hinges on useful software being available for those GPUs, and vendors can only do so much to enable and test this. If the DOE intends to build a portable, integrated AI/ModSim platform, it must enable key open-source projects to build and test on hardware architectures of interest. Otherwise, our ecosystem will lag behind the fast-moving AI ecosystem in industry. ECP developed tools to enable CI at HPC centers, but there is still work to do to ensure that these tools can be run securely and reliably to support a portable ecosystem for our community.

**Conclusion and Recommendations**

By supporting a cohesive, integrated portfolio of AI/ModSim scientific tools and libraries, DOE can facilitate more rapid development, greater collaboration across domains, and efficient scaling of AI/ModSim techniques to new scientific discoveries. We believe continued investments in community-driven ecosystems like the Extreme-scale Scientific Software Stack (E4S) and Spack can bridge the AI and ModSim communities by providing standardized, flexible, and performance-optimized environments that integrate AI and ModSim software.

Thank you for considering this input.

**Acknowledgment**

We sincerely thank all those who have contributed to E4S, Spack, and related topics in scientific software ecosystems. This material is based upon work supported by the U.S. Department of Energy, Office of Science, Office of Advanced Scientific Computing Research under contract numbers DE-AC02-06CH11357 (ANL), DE-AC52-07NA27344 (LLNL), and DE-NA0003525 (SNL), where Paratools is a subcontractor to ANL.

**Citations**

- Gamblin, Todd, et al. "Spack: A Flexible Package Manager for Supercomputers." *Proceedings of the International Conference for High Performance Computing, Networking, Storage and Analysis*, 2015, pp. 1-12. [Link to paper](#).
- Heroux, Michael A., et al. "E4S: Extreme-scale Scientific Software Stack." 2021. [E4S Website](#).
- Heroux, Michael A., et. al. "PESO: Partnering for Scientific-Software Ecosystem Opportunities." 2024. [PESO Project Website](#).